\documentstyle[11pt]{article}
\title{\vspace{4cm}\Large\bf \begin{center}
On the motion of rotating bodies
\end{center}
 \begin{center}
in  field gravity theory and general relativity
\end{center} }
 \author{
  \bf{Yurij Baryshev}
\\
 Institute of Astronomy, St.Petersburg State University,\\
 Staryj Peterhoff, 198504 St.Petersburg, Russia,\\
 e-mail: yuba@YB3255.spb.edu}

\date{~}

\begin{document}

\maketitle

\begin{abstract}
\noindent
On the basis of Lagrangian formalism of relativistic field theory
post-Newtonian equations of motion for a rotating body are derived
in the frame of Feynman's quantum field gravity theory (FGT)
and compared with
corresponding geodesic equations in general relativity (GR).
It is shown that in FGT the trajectory of a rotating test body
does not depend on a choice of a coordinate system. The equation
of translational motion of a gyroscope is applied to description
of laboratory experiments with free falling rotating bodies and
rotating bodies on a balance scale. Post-Newtonian relativistic
effect of periodical modulation of the
orbital motion of a rotating body is discussed
for the case of planets of the solar system and
for binary pulsars PSR B1913+16 and PSR B1259-63.
In the case of binary pulsars with known spin orientations
this effect gives a possibility to measure radiuses
of neutron stars.

\end{abstract}
\newpage
\section{Introduction}

The discovery and continuous observations of binary pulsars
has opened a new possibilities for testing relativistic gravity theories
\cite{tay}.
Timing observations of a pulsar in orbit with a stellar-mass companion
require  practical application of
different relativistic gravity effects such as
advance of periastron, gravitational redshift, time dilation,
bending of light, Lense-Thirring precession, geodetic precession,
and also gravitational radiation (see e.g. \cite{tay-wei} - \cite{wex2}).

All this effects are usually calculated in the frame of general
relativity (GR) which is the geometrical approach to gravity.
Observations of binary pulsars verified GR with high accuracy,
however it is still important to apply other gravity theories
for binary pulsars observations \cite{dam-esp, dam}.

Here it is considered the field gravity theory
which elements have been formulated
in 60-th by Feynman \cite{fey}, Thirring \cite{thir},
Kalman \cite{kalm} and others
(for recent discussion see Straumann \cite{stra}, Baryshev \cite{bar1}).
In {\em Lectures on Gravitation} \cite{fey}
Feynman gave a quantum relativistic field description of gravity
similar to other fundamental forces (strong, weak, electromagnetic).
Feynman's field gravity theory (FGT) is based on the quantum field theory
and presents an alternative {\em non-geometrical}
understanding the physics of gravitational interaction.
He emphasized that though the theory of gravitation
"has both a field and a geometrical interpretation" ,
"the geometrical interpretation is not really necessary or
essential to physics" (\cite{fey}, p.113).
According to Feynman, the gravity force between two masses
is caused by the exchange of gravitons
which are mediators of the
gravitational interaction and actually represent the quanta
of the relativistic tensor field $\psi^{ik}$ in Minkowski space
$\eta^{ik}$.

The fact that the field equations in FGT and GR is the same
in linear approximation and that in FGT
there is an iteration procedure
which saves gauge-invariance and leads to the same non-linear
field equations, is usually interpreted as complete identity
of GR and FGT. However, after the iteration is fulfilled
and geometrical interpretation is accepted,
the new puzzling conceptual problems  appear, among them the
spacetime horizons and the break down of the energy-momentum
conservation in geometrical approach
(see review by Straumann \cite{stra}).
Moreover the iteration procedure in the frame of FGT
is not uniquely defined operation because it requires
to fix an expression for the energy momentum tensor
of the gravitational field which is not uniquely defined
in Lagrangian formalism and needs additional physical
assumptions such as symmetry, positiveness of the
field energy, traceless  energy-momentum tensor
for massless fields (see Landau \& Lifshitz \cite{land-lif},
Bogolubov \& Shirkov \cite{bog-shir}, Baryshev \cite{bar1}).

All classical PN relativistic gravity effects are the same
in FGT and GR (see e.g. \cite{thir, bar1, bar3}). However
here it is shown that there is a post-Newtonian effect
which demonstrate a
difference between FGT and GR and which may be tested
by timing observations of binary pulsars and by future
laboratory experiments with rotating masses,
e.g. measuring a balance between differently rotating bodies.
This post-Newtonian effect leads to periodic modulation
of the orbital motion of a pulsar, caused by the
velocity dependence of gravity force acting on a rotating body.
In GR this effect is a coordinate-dependent concept and hence could not
be uniquely defined, while in FGT there is definite coordinate-independent
value for this relativistic gravity phenomenon.
The effect was firstly discussed in \cite{bar-sok}  as a prediction of
the field gravity theory for the Stanford gyroscope
 experiment \cite{ever}.
Here this effect is discussed in connection with binary pulsars
timing observations and  formulation of the
Galileo-like experiments with free falling rotating bodies
or measuring a balance between rotating bodies.

In sec.2  post-Newtonian  equations of motion
of relativistic test particles in relativistic tensor gravity field
are discussed and compared with geodesic equations of general relativity.
In sec.3  equations of motion for
a gyroscope orbiting around a central mass are derived
 and  applied to the cases
of laboratory experiments, planets of the solar system and
binary pulsars as  tests of the velocity dependence
of the gravity force acting on rotating bodies.
Conclusions are presented in sec.4.

\section{Post-Newtonian equations of motion for test particles
in field gravity theory and general relativity}

\subsection{Kalman's equation of motion in FGT}

The basic principles of FGT are the same as
for other relativistic quantum fields. These include the
Minkowski space, the fundamental role of the inertial frames,
the concepts of force and energy-momentum tensor, and also
quantum uncertainty principle and the many-path approach.

Following \cite{bar2} let us consider the motion of
relativistic test particle
with rest-mass $m_0$, 4-velocity $u^i$, 3-velocity $\vec{v}$
in the gravitational field described by the symmetric tensor
potential $\psi^{ik}$ on flat Minkowski space-time
where the Cartesian coordinates always exist
and the metric tensor $\eta^{ik}=diag(1,-1,-1,-1)$
(here we accept notations of the text-book \cite{land-lif}).

To derive the equation of motion in FGT
one may use the stationary action principle in the form
of the sum of the free particles and the interaction parts:
\begin{equation}
\label{delta1-S}
\delta S =\delta(\frac{1}{c}\int (\Lambda_{(p)} + \Lambda_{(int)})
d\Omega ) = 0
\end{equation}
where $d\Omega$ is the element of 4-volume and
the variation of the action is made with respect to the
particle trajectories $\delta x^i$
for {\em fixed} gravitational potential $\psi^{ik}$.

The free particle Lagrangian is
\begin{equation}
\label{Lambda-p}
\Lambda_{(p)} = -\eta_{ik} T_{(p)}^{ik}
\end{equation}
and the interaction Lagrangian is
\begin{equation}
\label{Lambda-int}
\Lambda_{(int)} = -\frac{1}{c^2}\psi_{ik} T_{(p)}^{ik}
\end{equation}
where the energy-momentum tensor (EMT) of the point particle is
\begin{equation}
\label{T-p-ik}
T_{(p)}^{ik}=m_0 c^2\delta(\vec{r}-\vec{r_p})\{1-\frac{v^2}{c^2}\}^{1/2}
u^i u^k
\end{equation}
Inserting equations (\ref{T-p-ik}),(\ref{Lambda-int})
into equation(\ref{delta1-S})
and taking into account that $ds^2=dx_ldx^l$ we get
\begin{equation}
\label{var-S}
\int (m_0c\delta(\sqrt{dx_ldx^l})+
\frac{m_0}{c}\delta(\psi_{ik}\frac{dx^idx^k}{\sqrt{dx_ldx^l}} )
= 0
\end{equation}
Opening the variation and integrating by parts, and taking into
account that the variation is made for the fixed values
of the integration limits,  we find
\begin{equation}
\label{var1-S}
\int {m_0cdu_i \delta dx^i +
\frac{2m_0}{c}d(u^k\psi_{ik})\delta x^i -
\frac{m_0}{c}d(\psi_{lk}u^l u^k u_i)\delta x^i -
\frac{m_0}{c}u^k \delta \psi_{ik} dx^i} = 0
\end{equation}
Consider also that
%
$$
du^i = \frac{du^i}{ds}ds ; \qquad  dx^i=u^ids ; \qquad
\delta \psi_{ki} = \psi_{ki,l}\delta x^l ;    \qquad
$$
$$
d(u^lu^ku_i\psi_{lk}) = u^lu_iu^k\psi_{lk,n}dx^n +
u^lu^k\psi_{lk}du_i + 2\psi_{lk}u^lu_idu^k ;
$$
finally we get the following equation of motion for test
particles in the field gravity theory [19]:
\begin{equation}
A_k^i \frac{dp^k}{ds}=-m_0 B^i_{kl}u^k u^l
\label{eq-motion}
\end{equation}
where $p^k=m_0cu^k$ is the 4-momentum of the test particle,
$(\cdot)_{,i} = \frac{\partial(\cdot)}{\partial x^i}$, and
\begin{equation}
A_k^i=(1- \frac{1}{c^2}\psi_{nl}u^n u^l)\eta_k^i -
\frac{2}{c^2}\psi_{kn}u^n u^i + \frac{2}{c^2}\psi_k^i
\label{Aik}
\end{equation}
\begin{equation}
B_{kl}^i= \frac{2}{c^2}\psi_{k,l}^i
-\frac{1}{c^2}\psi_{kl}^{\;\; ,i}
-\frac{1}{c^2}\psi_{kl,n}u^n u^i
\label{Bikl}
\end{equation}

The rest mass of the test particle can be canceled, hence in the
field gravity theory $m_{in}=m_g=m_0$ without initial equivalence
postulate as was emphasized by Thirring \cite{thir}.

The equation (\ref{eq-motion}) is identical to the
equation of motion derived by Kalman \cite{kalm} in another way,
by considering the relativistic Lagrange function
$L$ defined as $S=\int L \frac{ds}{c}$ and
relativistic Euler equation:
\begin{equation}
\frac{d}{ds}{(\frac{\partial L}{\partial u^k}u^k -L)u_i +
\frac{\partial L}{\partial u^i}} =
-\frac{\partial L}{\partial x^i}
\label{Kalman}
\end{equation}
Inserting in equation (\ref{Kalman})
the following expression of the relativistic Lagrange function
\begin{equation}
\label{Lagrange-function}
L=-m_0c^2 - m_0\psi_{ik}\frac{dx^i}{ds}\frac{dx^k}{ds}
\end{equation}
one gets Kalman's equations of motion \cite{kalm},
which may be transformed into equation (\ref{eq-motion}).


\subsection{PN equations of motion in FGT}

In post-Newtonian approximation we keep terms an order of
$v^2/c^2 \sim \varphi_N/c^2 \ll 1$ in equation (\ref{eq-motion}).
For PN accuracy we need calculations of the $\psi^{00}$
component with the same order, while other components of the
tensor gravitational potential $\psi^{ik}$ can be calculated
in Newtonian approximation.
Under these assumptions from equation (\ref{eq-motion})
for ($i=\alpha$) we get
the expression for the PN 3-dimensional
gravity force (which we shall call
the Poincar\'{e} gravity force remembering his pioneer work
in 1905 on relativistic gravity force in flat space-time):
\begin{eqnarray}
\vec{F}_{Poincare}= \frac{d\vec{p}}{dt} =
-m_0\{(1+\frac{3}{2}\frac{v^2}{c^2} +
3\frac{\phi}{c^2})\vec{\nabla}\phi
-3\frac{\vec{v}}{c}(\frac{\vec{v}}{c}
\cdot\vec{\nabla}\phi)\} \nonumber\\
-m_0\{3\frac{\vec{v}}{c}\frac{\partial \phi}{c\; \partial t} -
2\frac{\partial \vec{\Psi}}{c\; \partial t} +
2(\frac{\vec{v}}{c}\times rot\vec{\Psi})\}
\label{Poincare-PN-force}
\end{eqnarray}
where $\phi= \psi^{00}$, $\vec{\Psi}=\psi^{0\alpha} = -\psi_{0\alpha}$.

An important advance of FGT is that in  Minkowski space
according to Noether's theorem there is well-defined concept
of the energy-momentum tensor and its conservation in the form which is
usual for relativistic quantum fields. Hence the energy of gravity
field is localized positive physical quantity and is given by
the 00-component of the energy-momentum tensor (and not
pseudo-tensor as in GR). For post-Newtonian applications
it is sufficient to use the weak field approximation for the
calculation of the gravity field EMT,
which gives in the case of a static
spherically symmetric massive body the following value
for the 00-component (\cite{thir, bar3}):
\begin{equation}
T^{00}_{(g)} = \varepsilon_{(g)}=
\frac{1}{8\pi G}(\vec{\nabla}\varphi_N)^2 \;\;\;\frac{ergs}{cm^3}
\label{Tg00}
\end{equation}
where $\varphi_N$  is Newtonian potential. The energy
of the gravity field gives the following PN correction
to the 00-component of tensor gravitational potential
(\cite{thir, bar3}) :
\begin{equation}
\phi = \psi^{00} = \varphi_N +
\frac{1}{2}\frac{\varphi_N^2}{c^2}
\label{psi00}
\end{equation}
Hence in the frame of FGT the energy of the gravity field
is observable quantity which may be measured by observations
of test particles motion.
Take into account the expression (\ref{psi00}) for the
00-component of the gravitational potential, we get
corresponding PN 3-acceleration of the test particle :
$$
\frac{d\vec{v}}{dt} = -
(1+\frac{v^2}{c^2} +
4\frac{\varphi_N}{c^2})\vec{\nabla}\varphi_N +
4\frac{\vec{v}}{c}(\frac{\vec{v}}{c}\cdot\vec{\nabla}\varphi_N)
$$
\begin{equation}
+3\frac{\vec{v}}{c}\frac{\partial \varphi_N}{c\; \partial t} -
2\frac{\partial \vec{\Psi}}{c\;\partial t} +
2(\frac{\vec{v}}{c}\times rot\vec{\Psi})
\label{Poincare-PN-accel}
\end{equation}

>From the ($i=0$) component of equation (\ref{eq-motion})
it follows the expression for the work of the Poincar\'{e}
force:
\begin{equation}
\frac{dE_k}{dt}= \vec{v}\cdot\vec{F}_{Poincare}
=-m_0\vec{v}\cdot \{
(1-\frac{3}{2}\frac{v^2}{c^2} + 3\frac{\phi}{c^2})\vec{\nabla}\phi -
3\frac{\vec{v}}{c}\frac{\partial \phi}{c\;\partial t} +
2\frac{\partial \vec{\Psi}}{c\;\partial t}\}
\label{Poincare-work-PN}
\end{equation}

An important particular case is the static spherically symmetric
weak gravitational field for which $\vec{\Psi}=0$, $\partial
\phi/\partial t =0$, $\psi^{ik}=diag(\phi,\varphi_N,\varphi_N,\varphi_N)$
hence  PN 3-acceleration will have the
simple form:
\begin{equation}
(\frac{d\vec{v}}{dt})_{FGT} = -
(1+\frac{v^2}{c^2} +
4\frac{\varphi_N}{c^2})\vec{\nabla}\varphi_N +
4\frac{\vec{v}}{c}(\frac{\vec{v}}{c}\cdot\vec{\nabla}\varphi_N)
\label{FGT-accel}
\end{equation}
>From equation of motion (\ref{FGT-accel}) it is clear that the
acceleration of a test particle depends on the value and the
direction of its velocity and this is coordinate independent
relativistic gravity effect.

For circular motion $\vec{v}\perp \vec{\nabla}\varphi_N$
hence PN 3-acceleration is
\begin{equation}
(\frac{d\vec{v}}{dt})_{FGT} = -
(1+\frac{v^2}{c^2} +
4\frac{\varphi_N}{c^2})\vec{\nabla}\varphi_N
\label{FGT-accel-perp}
\end{equation}

For radial motion $\vec{v}\uparrow \downarrow \vec{\nabla}\varphi_N$
the 3-acceleration is
\begin{equation}
(\frac{d\vec{v}}{dt})_{FGT} = -
(1-3\frac{v^2}{c^2} +
4\frac{\varphi_N}{c^2})\vec{\nabla}\varphi_N
\label{FGT-accel-par}
\end{equation}
%


\subsection{PN equations of motion in GR}

In GR the equation of motion of a test particle is the
geodesic equation \cite{land-lif} :
\begin{equation}
\frac{du^i}{ds}=-\Gamma^i_{kl}u^k u^l
\label{geodesic-eq}
\end{equation}
where $u^i$ is 4-velocity of the test particle and
$\Gamma^i_{kl}$ is the Christoffel symbols.

Post-Newtonian geodesic equation has been carefully studied
in relativistic celestial mechanics and according to \cite{brum, soff}
PN 3-acceleration of a test particle in the case of static
spherically symmetric gravitational field is:
$$
(\frac{d\vec{v}}{dt})_{GR} = -
\{1+(1+\alpha)\frac{v^2}{c^2} +
(4-2\alpha)\frac{\varphi_N}{c^2}
-3\alpha(\frac{\vec{r}}{r}\cdot\frac{\vec{v}}{c})^2
\}\vec{\nabla}\varphi_N
$$
\begin{equation}
+(4-2\alpha)\frac{\vec{v}}{c}(\frac{\vec{v}}{c}\cdot\vec{\nabla}\varphi_N)
\label{GR-accel}
\end{equation}
where $\vec{v}=d\vec{r}/dt$, $\varphi_N=-GM/r$,
$\vec{\nabla}\varphi_N=GM\vec{r}/r^3$.

The very important quantity in equation(\ref{GR-accel}) is
the parameter of coordinate system $\alpha$ which
depend on the choice of particular coordinate system
and for example has the following values:
$\alpha = 2$ for Painlev\'{e} coordinates,
$\alpha = 1$ for Schwarzschild coordinates,
$\alpha = 0$ for harmonic or isotropic coordinates.
Hence the trajectory of a test particle is a coordinate-dependent
concept, i.e. it depends on
the choice of a coordinate system due to parameter $\alpha$.

In the case of circular motion the 3-acceleration is:
\begin{equation}
(\frac{d\vec{v}}{dt})_{GR} = -
\{1+(1+\alpha)\frac{v^2}{c^2} +
(4-2\alpha)\frac{\varphi_N}{c^2}
\}\vec{\nabla}\varphi_N
\label{GR-accel-perp}
\end{equation}

For radial motion the 3-acceleration is:
\begin{equation}
(\frac{d\vec{v}}{dt})_{GR} = -
\{1-3\frac{v^2}{c^2} +
(4-2\alpha)\frac{\varphi_N}{c^2}
\}\vec{\nabla}\varphi_N
\label{GR-accel-par}
\end{equation}
According to \cite{brum} in GR there are "coordinate-dependent
(unmeasurable) quantities" and "measurable quantities" which
may be directly obtainable from observation without involving
any theoretical data (such as the laws of light propagation).
For example the radius $r$ of the circular orbit is a
coordinate-dependent, unmeasurable quantity, while the
masses $M$ and $m$ do not depend on coordinate system
and hence are measurable quantities.

The geodesic equations (\ref{GR-accel},\ref{GR-accel-perp},
\ref{GR-accel-par}) are identical with equations
(\ref{FGT-accel},\ref{FGT-accel-perp},\ref{FGT-accel-par})
of FGT only for harmonic or isotropic coordinates where
$\alpha =0$. Fock \cite{fock}  also emphasized the special role
of harmonic coordinates in GR.

\section{Translational motion of a rotating body in FGT}

\subsection{PN equations of motion of a gyroscope in FGT}

Let us consider Poincar\'{e} gravity force acting on a rotating body.
>From equation (\ref{Poincare-PN-force}) in the case of a gyroscope
motion in static spherically symmetric gravitational field
it follows the expression for the elementary gravity force $dF_P$
acting on each elementary mass $dm$ of the gyroscope:
\begin{equation}
d\vec{F}_P = -\{
(1+\frac{3}{2}\frac{v^2}{c^2} +
4\frac{\varphi_N}{c^2})\vec{\nabla}\varphi_N -
3\frac{\vec{v}}{c}(\frac{\vec{v}}{c}\cdot\vec{\nabla}\varphi_N)\}dm
\label{dm-force}
\end{equation}
For a rotating solid body the total gravity force is the sum of
elementary forces acting on elementary masses:
\begin{equation}
\vec{F}_P =\int d\vec{F}_P
\label{total-force}
\end{equation}
Taking into account that the velocity $v$ of an element $dm$
may be presented in the form
\begin{equation}
v = \vec{V} + [\vec{\omega}\vec{r}]
\label{v}
\end{equation}
where $\vec{V}$ is the translational velocity of the body,
$\vec{\omega}$ is the angular velocity, $\vec{r}$ is the radius
vector of an element $dm$ relative to its  center of inertia,
so that $\int \vec{r}dm=0$.

Inserting (\ref{v}) into (\ref{dm-force}) and (\ref{total-force})
we get
\begin{eqnarray}
\vec{F}_P = -M\{
(1+\frac{3}{2}\frac{V^2}{c^2} +
4\frac{\varphi_N}{c^2}+
\frac{3}{2}\frac{I\omega ^2}{Mc^2}
)\vec{\nabla}\varphi_N        \nonumber\\
-3\frac{\vec{V}}{c}(\frac{\vec{V}}{c}\cdot\vec{\nabla}\varphi_N) -
\frac{3}{Mc^2}\int [\vec{\omega}\vec{r}]
([\vec{\omega}\vec{r}]\cdot \vec{\nabla}\varphi_N) dm \}
\label{gyroscope-force}
\end{eqnarray}
where $M=\int dm$ is the total mass of the body,
$I$ is its moment of inertia.

Under the gravity force (\ref{gyroscope-force}) the rotating body
will get the 3-acceleration according to general relation \cite{land-lif} :
\begin{equation}
\frac{dv}{dt}=\frac{\sqrt{1-v^2/c^2}}{m_0}
(\vec{F} - \frac{\vec{v}}{c}(\frac{\vec{v}}{c}\cdot \vec{F}))
\label{accel-force}
\end{equation}
which may be written in the form:
\begin{eqnarray}
\frac{d\vec{V}}{dt} = -
(1+\frac{V^2}{c^2} +
4\frac{\varphi_N}{c^2}+
\frac{3}{2}\frac{I\omega ^2}{Mc^2}
)\vec{\nabla}\varphi_N     \nonumber\\
+4\frac{\vec{V}}{c}(\frac{\vec{V}}{c}\cdot\vec{\nabla}\varphi_N) +
\frac{3}{Mc^2}\int [\vec{\omega}\vec{r}]
([\vec{\omega}\vec{r}]\cdot \vec{\nabla}\varphi_N) dm
\label{gyroscope-accel}
\end{eqnarray}
Equation of motion of a rotating body (\ref{gyroscope-accel})
shows that the orbital velocity of the center of mass of the body
will have additional perturbations due to rotation of the body.
The last term in (\ref{gyroscope-accel}) depends on
direction and value of the angular velocity $\vec{\omega}$.
It has an order of magnitude $v_{rot}^2/c^2$ and may be measured
in laboratory experiments and astrophysical observations.

\subsection{Laboratory experiments with rotating bodies}

The most straightforward application of equation (\ref{gyroscope-accel})
is to perform "Galileo-2000" experiment (which is an
improved version of famous Stevinus-Grotius-Galileo experiment
with free falling bodies in the Earth gravity field) just
taking into account rotation of the bodies.

Indeed let us consider three balls on the top of a tower
(like the 110-m Drop Tower of the Bremen University).
The first ball is non-rotating and according to equation
(\ref{gyroscope-accel})  its free fall acceleration is:
\begin{eqnarray}
\vec{g}_1=
(\frac{d\vec{V}}{dt})_1 = -
(1-3\frac{V^2}{c^2} +
4\frac{\varphi_N}{c^2}
)\vec{\nabla}\varphi_N
\label{g1}
\end{eqnarray}

Let the rotation axis of the second ball be parallel to
the gravity force, i.e. $\vec{\omega}\|\vec{\nabla}\varphi_N$,
hence its free fall acceleration is:
\begin{eqnarray}
\vec{g}_2=\vec{g}_1
(1+\frac{3}{5}\frac{R^2\omega^2}{c^2} )
\label{g2}
\end{eqnarray}
where it is taken into account that for homogeneous ball
with radius $R$ and mass $M$
the moment of inertia is $I=\frac{2}{5}MR^2$.

Let the rotation axis of the third ball be orthogonal to the gravity
force, i.e. $\vec{\omega}\perp\vec{\nabla}\varphi_N$,
hence its free fall acceleration is:
\begin{eqnarray}
\vec{g}_3=\vec{g}_1
(1-\frac{3}{10}\frac{R^2\omega^2}{c^2})
\label{g3}
\end{eqnarray}

>From equations (\ref{g1}),(\ref{g2}),(\ref{g3}) it follows that
considered three balls will reach the ground at different
moments. True the difference is very small, for example if the
radius of the ball is $R=10$ cm and its angular velocity
$\omega =10^3$ rad/sec, then the expected difference in
falling time from 110 m tower
will be $\Delta t \approx (1/2)(\Delta g/g)t
\approx 2.5\times 10^{-13}$ sec.

For the NASA's gravity probe B (GPB) experiment \cite{ever} the radius
of the gyroscope is $R\approx 2$ cm and it rotates with about
2000 revolutions per second. The expected effect of the periodical
perturbation of the acceleration of the gyroscope center of mass
relative to non-rotating satellite center of mass is about
$\delta g/g \approx 6\times 10^{-13}$ which is in principle
measurable \cite{bar-sok}.

Another type of laboratory experiment for direct testing the velocity
dependence of the Poincar\'{e} gravity force
(\ref{gyroscope-force}) is to weigh a rotating
bodies. If two bodies are at the balance and at a moment they
start to rotate with different orientations of the rotation axes
then the balance will be violated and hence measured by a scale.
The expected difference in forces is again  about $\omega^2 R^2/c^2$.

In these laboratory experiments there are no problem
with choosing a coordinate system at all. The height of a tower
and the moments of the contact of the rotating bodies with
the ground, and also readings of a balance scales
are directly measurable quantities. Equation of motion
(\ref{FGT-accel}) in field gravity theory gives the uniquely
defined value for these laboratory experiments, while
geodesic equation (\ref{GR-accel}) includes arbitrary parameter
$\alpha$ of a coordinate system.


\subsection{Periodical modulation of planet's orbits}

Post-Newtonian celestial mechanics in FGT
does not depend on a choice of coordinate systems and
is identical with PN approximation of GR only in
harmonic or isotropic coordinates. This is why all
classical post-Newtonian relativistic gravity effects
are the same in FGT and GR. Note that according to \cite{mtw}
"astronomers have adopted the fairly standard convention
of using "isotropic coordinates" rather than "Schwarzschild
coordinates" when analyzing the solar system"(\cite{mtw},p.1097).

According to \cite{brum}  in GR the effects which depend on the choice
of coordinate system (e.g. the parameter $\alpha$) are
unmeasurable. However there are such effects which are
coordinate dependent in GR but are coordinate independent
in FGT. An example of such effect is additional perturbation
terms in equations of motion of rotating bodies which
are determined by angular velocities of the bodies
(see equation (\ref{gyroscope-accel})).

For a planet rotating with the angular velocity $\vec{\omega}$ in
orbit with semi-malor axis $a$,  and
orbital angular momentum $\vec{L}$ the effect of rotation
leads to periodical modulation of the orbital radius.
The amplitude of the radius perturbation $\Delta a$
may be roughly estimated as:
\begin{equation}
\Delta a \approx a \frac{\omega^2 R^2}{c^2} \sin^2 \theta
\label{delta-a}
\end{equation}
where $R$ is the radius of planet, $\theta$ is the angle between
$\vec{\omega}$ and $\vec{L}$.

According to equation (\ref{delta-a}) the amplitude of the orbital
modulation is about 6 cm for the Earth, 1 cm for  Mars,
4.2 m for Jupiter and 340 m for Saturn.
Future high accuracy radar observations of the solar system
will test this tiny effects.

\subsection{Rapidly rotating stars and pulsars in binary systems}

In a binary stellar system the rapidly rotating companion will produce
the relativistic effect of the periodical modulation of
the orbital motion. However for usual
stars there are many other non-relativistic effects which usually
hide  small relativistic effects.

The best candidates for testing relativistic gravity effects
are pulsars in binary systems. Timing observations of pulsars,
i.e. the measurements of the time of arrival of pulsar signals
at a radio telescope, are the high-precision experiments in
astronomy. About 10$\%$ of the known pulsars are members of binary
star systems, i.e. in orbit around a white dwarf, neutron star,
or main-sequence star companion. Timing observations of these
binary pulsars may be used for testing the periodical modulation
of the orbit of the rotating bodies.

In the case of equal masses of stars it is needed to generalize
the above test-body approach to the case of two-body problem.
However for rough estimation of the expected effect it is enough
to use equation (\ref{gyroscope-accel}).
According to  (\ref{delta-a})
the amplitude of the modulation of the orbit
are determined  by the spin angular velocity $\omega$, the radius $R$
of the pulsar (or more exactly, the distribution
of the mass and velocity in the pulsar, i.e. its moment of inertia
$I_{\alpha , \beta}$ and spin angular momentum $\vec{S}$),
semi-major axis $a$ and the angle $\theta_p$ between pulsar's spin
and direction of orbital angular momentum.

Because the amplitude of the modulation effect is proportional
to the semi-major axis, the largest effect may be observed
for the special class of the long-orbital period binary pulsars \cite{wex3}.
For known $\omega = 2\pi /P_p$ , $a$ and $R$ it is possible
to determine the angle $\theta_p$.
If for a pulsar there is sufficiently accurate
estimation of the angle $\theta_p$,
then it is possible to estimate the radius
of the pulsar $R$.

Let us make estimation for the expected effect in the case of
two well-studied binary pulsars PSR B1913+16 and PSR B1259-63.
The amplitude of the orbital modulation is given by (\ref{delta-a})
and the value of the expected variation in arrival times of pulses
$\Delta \tau = \Delta a /c$ may be written in the form:
\begin{equation}
\Delta \tau \approx 1.75\cdot 10^{-5}(sec) (\frac{x}{1\; s})
(\frac{50\; ms}{P_p})^2 (\frac{R}{10\; km})^2
\sin^2 \theta_p
\label{delta-tau}
\end{equation}
where $x= a\cdot sin\;i/c$ is the projected semi-major axis,
$i$ is the inclination angle of the orbit,
$P_p$ is the period of the pulsar,
$R$ is the radius of the neutron star.
and $\theta_p$ is the misalignment angle of the pulsar spin
to the orbital momentum vector.

For PSR B1913+16: period $P_p=59$ ms, $x=2.34$ s, and $\theta_p
\approx 22^{\circ}$ \cite{kram}, hence the expected periodical variations
in arrival time due to rotational effect is about 6 $\mu$s.
For PSR B1259-63 we have $P_p=47.8$ ms, $x=1296$ s \cite{wex3} and for
$\theta =10^{\circ}$ the time variations will be about 0.7 ms.
These numbers are close to the time residuals existing for these
pulsars and it shows possibility to test the discussed effect.

\section{Conclusions}

Using Lagrangian formalism of relativistic field theory
it is derived  PN equations of motion of a rotating body
in the framework of Feynman's field gravity theory.

The trajectory of a test particle
does not depend on a choice of a coordinate system
and this is an essential property of field gravity theory
which allow to make uniquely defined predictions for
terrestrial laboratory experiments with rotating masses
in Earth's gravity field.
In these laboratory experiments there are no problem
with choosing a coordinate system at all. The height of a tower
and the moments of the contact of the rotating bodies with
the ground, and also readings of a balance scales
are directly measurable quantities. Equation of motion
(\ref{FGT-accel}) in field gravity theory gives the uniquely
defined value for these laboratory experiments, while
geodesic equation (\ref{GR-accel}) includes arbitrary parameter
$\alpha$ of a coordinate system.

In the case of astronomical
observations of distant objects it is important to
consider also effects of emission, propagation and
detection of photons, however it does not change
the concept of coordinate independence of the trajectory
of a test particle.

Post-Newtonian  periodical
modulation of planetary motion in the solar system
may be measured by future radar observations.
Timing observations of pulsars in binary systems
is the best test of this relativistic gravity effect.
The expected arrival time variation due to pulsar rotation
lies in interval from 1 $\mu$s up to 1 ms depending mainly on
the value of the semi-major axis of pulsar orbit,
misalingnment angle of pulsar spin, angular velocity of its rotation
and the radius of a neutron star.
In the case of binary pulsars with known spin orientations
this effect gives a possibility to measure radiuses
of neutron stars.

\end{document}